\documentclass[twocolumn,showpacs,preprintnumbers,amsmath,amssymb,superscriptaddress,prb]{revtex4}
\usepackage{graphicx}% Include figure files
\usepackage{dcolumn}% Align table columns on decimal point
\usepackage{bm}% bold math

\usepackage{color}
\definecolor{red}{rgb}{0,0,0}

\definecolor{blue}{rgb}{0,0,0}

\definecolor{green}{rgb}{0,0,0}

\begin{document}
\preprint{APS}

\title{Incommensurate spin-density wave and multiband superconductivity in Na$_{x}$FeAs as revealed by nuclear magnetic resonance}

\author{M. Klanj\v{s}ek}
\affiliation{Jo\v{z}ef Stefan Institute, Jamova 39, 1000 Ljubljana, Slovenia}

\author{P. Jegli\v{c}}
\affiliation{Jo\v{z}ef Stefan Institute, Jamova 39, 1000 Ljubljana, Slovenia}

\author{B. Lv}
\affiliation{Department of Chemistry and TCSUH, University of Houston, Houston, TX 77204-5002, USA}
\affiliation{Texas Center for Superconductivity, University of Houston, Houston, TX 77204-5002, USA}

\author{A. M. Guloy}
\affiliation{Department of Chemistry and TCSUH, University of Houston, Houston, TX 77204-5002, USA}

\author{C. W. Chu}
\affiliation{Texas Center for Superconductivity, University of Houston, Houston, TX 77204-5002, USA}

\author{D. Ar\v{c}on}
\email{denis.arcon@ijs.si}
\affiliation{Jo\v{z}ef Stefan Institute, Jamova 39, 1000 Ljubljana, Slovenia}
\affiliation{Faculty of Mathematics and Physics, University of Ljubljana, Jadranska 19, 1000 Ljubljana, Slovenia}

\date{\today}

\begin{abstract}

We report a $^{23}$Na and $^{75}$As nuclear magnetic resonance (NMR) investigation of Na$_{x}$FeAs series ($x=1$, $0.9$, $0.8$) exhibiting a spin-density wave (SDW) order below $T_{\rm SDW}=45$, $50$ and $43$~K for $x=1$, $0.9$, $0.8$, respectively, and a bulk superconductivity below $T_c\approx 12$~K for $x=0.9$. Below $T_{\rm SDW}$, a spin-lattice relaxation reveals the presence of gapless particle-hole excitations in the whole $x$ range, meaning that a portion of the Fermi surface remains gapless. The superconducting fraction as deduced from the bulk susceptibility scales with this portion, while the SDW order parameter as deduced from the NMR linewidth scales inversely with it. The NMR lineshape can only be reproduced assuming an incommensurate (IC) SDW. These findings qualitatively correspond to the mean-field models of competing interband magnetism and intraband superconductivity, which lead to an IC SDW order coexisting with superconductivity in part of the phase diagram.

\end{abstract}

\maketitle

\section{Introduction}

Iron pnictides emerged recently as a new class of materials with relatively high superconducting (SC) critical temperatures, $T_c$, surpassed only by cuprates.\cite{Paglione_2010, Johnston_2010, Wilson_2010} Since electron-phonon coupling strength is too weak to account for $T_c$ as high as 55~K,\cite{Haule_2008, Boeri_2008} unconventional pairing mechanisms are vividly discussed in the literature.\cite{Paglione_2010, Johnston_2010, Wilson_2010} In contrast to other high-$T_c$ families,\cite{Lee_2006, Chu_2009, Cs3Science, Ganin2010} Coulomb correlations are believed to be only weakly to moderately strong. However, similarly as in other high-$T_c$ families, a SC state appears next to a spin-density wave (SDW) state suggesting that the spin fluctuations play an important role in the mechanism of SC pairing.\cite{Luetkens_2009, Chen_2009} This notion is strongly supported by, e.g., nuclear magnetic resonance (NMR) experiments,\cite{Ishida_2009, Ning_2010} which provide a direct evidence for the  spin fluctuations that even get enhanced close to $T_c$. The way carrier doping affects spin fluctuations and stabilizes superconductivity as the ground state is presently still debated.  

The members of a 111 family ($A$FeAs where $A$ is Li or Na)~\cite{Tapp_2008, Chu_2009_2, Sasmal_2009} exhibit the structure, which is a simplified version of the structure of 1111 and 122 iron pnictides:\cite{Paglione_2010, Johnston_2010, Wilson_2010} FeAs layers comprised of edge-sharing FeAs$_4$ tetrahedra are separated by double layers of Li or Na atoms. LiFeAs and NaFeAs are thus expected to be electronically very similar to each other and also to other layered iron pnictides. However, undoped LiFeAs and NaFeAs adopt different ground states. LiFeAs is a bulk superconductor with $T_c=18$~K,\cite{Tapp_2008} while according to the muon-spin rotation, neutron diffraction and NMR studies NaFeAs exhibits a SDW order below $T_{\rm SDW}\approx 40$~K, which is essentially of the same type as found in LaFeAsO but with a reduced order parameter $0.1-0.2\mu_B$ (Ref.~\onlinecite{Parker_2009}), $0.09\mu_B$ (Ref.~\onlinecite{Li_2009}) and $0.3\mu_B$ (Ref.~\onlinecite{Yu_2010}), respectively. Bulk superconductivity is reported for the Na-deficient samples Na$_x$FeAs in a narrow region around $x=0.9$ (Ref.~\onlinecite{Sasmal_2009}), and for the Co and Ni doped samples NaFe$_{1-x}$Co$_x$As and NaFe$_{1-x}$Ni$_x$As.\cite{Parker_2010} These differences open a number of important questions, which are not only relevant for the 111 family but also for iron pnictides in general: (i) How does the SDW order respond to carrier doping? (ii) Why is the SDW order parameter so small and how is it related to the details of the multiband electronic structure? (iii) Do SDW and bulk SC phases coexist and how do important physical parameters, such as $T_{\rm SDW}$, the order parameter, and $T_c$ behave in the critical doping regime?  

To tackle the above issues, we perform a systematic $^{23}$Na and $^{75}$As NMR investigation of Na$_{x}$FeAs series with $x=1$, $0.9$, $0.8$. We confirm a bulk superconductivity below $T_c\approx 12$~K only for the optimal doping $x=0.9$. Quite surprisingly, we find that the SDW order is incommensurate (IC) and present in the whole $x$ range, the most prominent effect being a severe suppression of the SDW order parameter at the optimal doping. Meanwhile, we find that a portion of the Fermi surface remains gapless below $T_{\rm SDW}$ in the whole $x$ range. This portion is the biggest at the optimal doping, where we observe the coexistence of the SC order and the IC SDW order on atomic scale below $T_c$.

\section{Experiment and discussion}

\begin{figure}[t]
\includegraphics[width=1\linewidth]{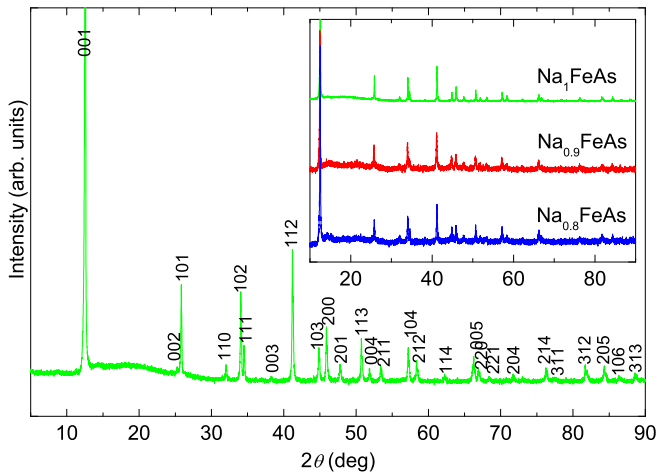}
\caption{Room temperature XRD of Na$_1$FeAs sample. All peaks can be indexed within $P4/nmm$ space group showing that the sample does not contain any impurity phases. Inset: Comparison of Na$_x$FeAs ($x=1$, $0.9$, $0.8$) XRDs showing that the three samples are isostructural.}
\label{XRD}
\end{figure}

Na$_{x}$FeAs ($x=1$, $0.9$, $0.8$) powdered samples were synthesized as described elsewhere.\cite{Sasmal_2009, Parker_2009, Chu_2009_2} X-ray phase pure samples of FeAs powder were first synthesized from pure elements of Fe (pieces, 99.99\%) and As (lumps, 99.999\%) in sealed quartz containers at 600-800~$^\circ$C. The ternary Na$_x$FeAs samples ($x=1$, $0.9$, $0.8$) were then prepared from stoichiometric amounts of the starting materials of Na (ingot, 99.95\%) with FeAs in sealed Nb tubes under Ar (which are subsequently sealed in quartz tubes under vacuum) at 750~$^\circ$C. All the preparative manipulations were carried out in a purified argon atmosphere glove box with total O$_2$ and H$_2$O level $<1$~ppm. The quality and composition of the samples were checked prior to the NMR experiments by powder X-ray diffraction (XRD) using the Panalytical Xpert Diffractometer. All peaks of the Na$_1$FeAs pattern match the space group $P4/nmm$ (Fig.~\ref{XRD}). The lattice parameters for the Na$_1$FeAs sample were refined as $a=3.9520(2)$~${\rm\AA}$ and $c=7.0461(5)$~${\rm\AA}$, in perfect agreement with the published data.\cite{Sasmal_2009, Parker_2009, Chu_2009_2, Li_2009} No additional peaks that would indicate the presence of impurity phases were found. The Na$_{0.9}$FeAs and Na$_{0.8}$FeAs samples were found isostructural to Na$_1$FeAs (inset to Fig.~\ref{XRD}) and also their XRDs do not indicate any impurities.

The bulk dc magnetic susceptibility of the Na$_x$FeAs powders was measured with a commercial Quantum Design MPMS system in a small magnetic field of $10$~Oe under zero-field-cooling conditions. The dc magnetization exhibits characteristic superconducting diamagnetic behavior below $T_c\approx 12$~K [Fig.~\ref{fig2}(a)] for all three samples. However, in contrast to the weak doping dependence of $T_c$, the superconducting fractions strongly vary with $x$. The bulk superconductivity (superconducting fraction close to $100\%$) is found only in the $x=0.9$ sample, while small superconducting fractions of about $20\%$ are found in the $x=1$ and $0.8$ samples.

\begin{figure}[t]
\includegraphics[width=0.9\linewidth]{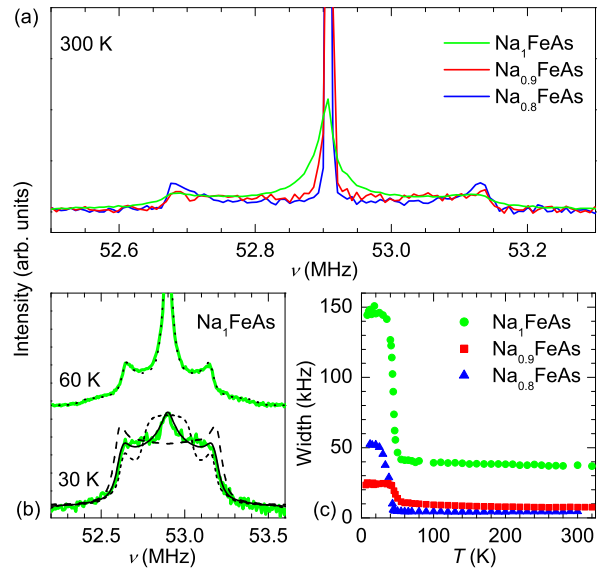}
\caption {(a) $^{23}$Na NMR spectra of powdered Na$_x$FeAs samples ($x=1$, $0.9$, $0.8$) at $300$~K. (b) $^{23}$Na NMR spectra of Na$_1$FeAs at $60$~K and $30$~K (thick green lines). Comparison is made to the simulated spectra for a powdered sample with $\nu_Q=530$~kHz and with the following type of the SDW order: no order (thin dotted line), IC SDW with the order parameter $0.28\mu_B$ (thin solid line), commensurate SDW with the order parameter $0.28\mu_B$ (thin dashed line) and commensurate SDW with the order parameter $0.12\mu_B$ (thin short dashed line). (c) Temperature dependence of the width of $^{23}$Na central transition in the three samples.}
\label{fig1}
\end{figure}

For the NMR experiments we sealed the samples into the $5$~mm pyrex tubes under dynamic vacuum conditions in order to avoid possible contamination with air. In Fig.~\ref{fig1}(a) we compare the powder $^{23}$Na (spin $I=3/2$) NMR spectra measured in Na$_{x}$FeAs samples for different $x$ at $300$~K. The spectra were recorded in a magnetic field of $4.7$~T by frequency sweeping. The $^{23}$Na Larmor frequency $52.903~$MHz was determined using the NaCl standard. All spectra exhibit a typical quadrupole powder pattern defined with the quadrupole splitting $\nu_Q=460$~kHz and the electric field gradient (EFG) asymmetry parameter $\eta=0$, in agreement with the Na site symmetry.\cite{Li_2009} Symmetric singularities belonging to the satellite transitions ($-3/2\leftrightarrow -1/2$ and $1/2\leftrightarrow 3/2$) are spaced by $\nu_Q$. In all cases, the central transition ($-1/2\leftrightarrow 1/2$) is located almost at the Larmor frequency meaning that the contact hyperfine shift is negligible. This indicates that the residual spin density at the Na site is very small implying almost complete charge transfer to the electronically active FeAs layer. Na vacancies in Na-deficient samples are expected to result in the local disorder and thus broadened NMR lines. In contrast, the measured $^{23}$Na central transition linewidth in the temperature range $~50-300$~K is the biggest for the stochiometric composition $x=1$. Although we do not completely understand this, it may imply that the migration of the Na$^+$ ions is enhanced in more Na-deficient samples, and it averages out the local field inhomogeneities resulting in the narrowed NMR lines.

The central $^{23}$Na transition undergoes a steep broadening below $T_{\rm SDW}=45$, $50$ and $43$~K for $x=1$, $0.9$, $0.8$, respectively, as demonstrated in Fig.~\ref{fig1}(b) for $x=1$. The $^{23}$Na linewidth increase is a direct measure of the local magnetic field distribution at the Na site and is thus proportional to the SDW order parameter. As shown in Fig.~\ref{fig1}(c), the linewidth saturates to quite different values for different $x$ at low temperature. This indicates that the saturated SDW order parameter exhibits a pronounced variation with $x$, although $T_N$ depends only mildly on $x$. We stress that this cannot be due to some phase segregation because our $^{23}$Na NMR spectra are characterized by a single pair of quadrupole parameters $\nu_Q$ and $\eta$ and thus suggest the presence of a single phase in all samples, in compliance with our XRD data.

Fig.~\ref{fig2}(b) shows the temperature dependence of $1/^{23}T_1T$, where $^{23}T_1$ is the $^{23}$Na nuclear spin-lattice relaxation time. To determine $^{23}T_1$ we used an inversion recovery method and fitted the spin-lattice relaxation data to the theoretical recovery curve for spin $I=3/2$ nuclei: $M(t)-M_0\propto 0.1\exp[-(t/T_1)^\gamma]+0.9\exp[-(6t/T_1)^\gamma]$. We allowed for the stretching exponent $\gamma$ as spin fluctuations in the layered Na$_x$FeAs are expected to be anisotropic, resulting in the distribution of $^{23}T_1$ in powdered samples. In addition, any disorder in the sample is expected to broaden the distribution of $^{23}T_1$, resulting in a lower $\gamma$. As shown in Fig.~\ref{fig2}(c), $\gamma$ is temperature independent above $100$~K and takes basically the {\em same value} $\sim 0.74$ for all three samples, suggesting that the degree of disorder in the samples is comparable. For all $x$ the temperature dependence of $1/^{23}T_1T$ strongly deviates from the simple Korringa relation. In particular, above $100$~K the data nicely follow an empirical power law $1/^{23}T_1T=aT^\alpha$, where the exponent $\alpha$ depends on the doping level $x$. Although the rising dependence of $1/T_1T$ on temperature is frequently ascribed to the pseudo-gap behavior, we suggest that it may be a direct consequence of the particular shape of the density of electronic states in the vicinity of Fermi level. In fact, taking the density of states as calculated for NaFeAs in Ref.~\onlinecite{Jishi_2010} and using the general expression for the nuclear spin-lattice relaxation in metals,\cite{Abragam} we are able to roughly reproduce the power law dependence $1/^{23}T_1T=aT^\alpha$, where $\alpha$ depends on the position of the Fermi energy in the density of states. Below $100$~K, the $1/^{23}T_1T$ data exhibit an upturn for all $x$ before reaching the maximum at $T_{\rm SDW}$, implying a growing importance of spin fluctuations close to $T_{\rm SDW}$. For their contribution to the spin-lattice relaxation we assume the form ${B/(T-\theta)^\beta}$, where $\theta$ is the Curie-Weiss temperature and $\beta=0.5$ as expected for the two-dimensional itinerant antiferromagnets.\cite{Nakai} Our model for the paramagnetic state thus consists of two terms, 
\begin{equation}
\frac{1}{^{23}T_1T}=aT^\alpha+\frac{B}{(T-\theta)^\beta},
\label{T1high}
\end{equation}
corresponding to the intraband and interband electronic scattering processes, respectively. As shown in Fig.~\ref{fig2}(a), the model fits the spin-lattice relaxation data very well. Fitting parameters are summarized in Table~\ref{table}. We find a surprisingly weak doping dependence of the parameters $B$ and $\theta$ related to interband processes, while the parameters $\alpha$ and $a$ related to intraband processes depend strongly on the doping level. Although $\theta$ is the smallest for the optimally doped sample Na$_{0.9}$FeAs, it does not approach zero as in other iron pnictides, e.g., in Co-doped BaFe$_2$As$_2$.\cite{Ning_2010}

\begin{table}[b]
\caption{Parameters obtained in fitting the temperature dependence of the $^{23}$Na ($^{7}$Li) spin-lattice relaxation rate from Fig.~\ref{fig2}(b) in Na$_{x}$FeAs (Li$_1$FeAs) samples to Eq.~(\ref{T1high}).  
\label{tab1}}
\begin{ruledtabular}
\begin{tabular}{cccccc}
& Sample & $a$  & $\alpha$ & $B$  & $\theta$  \\
& & ($10^{-5}$ s$^{-1}$K$^{-1-\alpha}$) & & (s$^{-1}$K$^{-0.5}$) & (K)\\
\hline
\hline
$^{23}$Na & Na$_1$FeAs  & 0.5 & 1.26 & 0.0038 & 42\\
& Na$_{0.9}$FeAs  & 3.3 & 0.93 & 0.0065 & 37\\
& Na$_{0.8}$FeAs & 0.13 & 1.57 & 0.0075 & 39\\
$^{7}$Li & Li$_1$FeAs & 75 & 0.54 & 0.056 & -3\\
\end{tabular}
\end{ruledtabular}
\label{table}
\end{table}

\begin{figure}[t]
\includegraphics[width=0.75\linewidth]{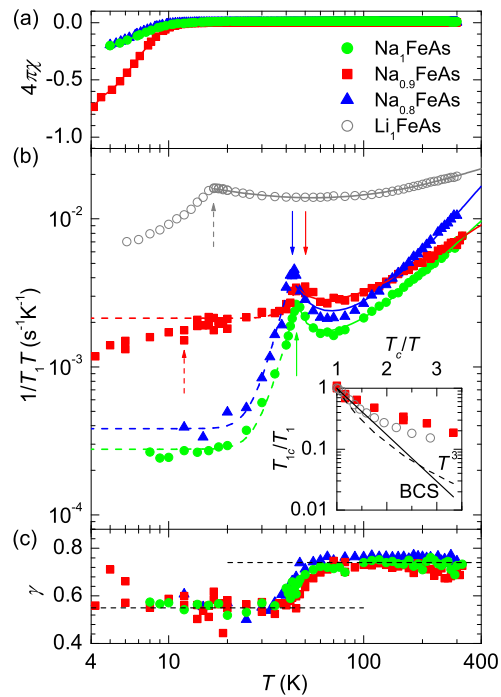}
\caption{(a) Temperature dependence of the dc spin susceptibility in Na$_x$FeAs samples for $x=1$, $0.9$, $0.8$. (b) Temperature dependence of $^{23}$Na $1/T_1T$ measured in Na$_x$FeAs samples. The $^{7}$Li data measured in Li$_1$FeAs are added for comparison. Arrows indicate the SC (dashed) and the SDW (solid) transition temperatures. Solid and dashed lines are fits to Eqs.~(\ref{T1high}) and (\ref{T1low}), respectively. Inset shows the dependence of $^{23}$Na $T_1(T_c)/T_1$ on $T_c/T$. The rate of suppression is significantly smaller than BCS or even unconventional $T^3$ behavior. (c) Temperature dependence of the stretching exponent $\gamma$ used in extracting the $^{23}T_1$ values from the spin-lattice relaxation data. Dashed lines indicate the temperature independent values $\sim 0.55$ and $\sim 0.74$ adopted by all three Na$_x$FeAs samples below and above the corresponding $T_{\rm SDW}$, respectively.}
\label{fig2}
\end{figure}

As all five Fe $d$-derived bands cross the Fermi level, we investigate now how different parts of the Fermi surface evolve below $T_{\rm SDW}$ upon doping. The $1/^{23}T_1T$ data below $T_{\rm SDW}$ exhibit a strong temperature and doping dependence [Fig.~\ref{fig2}(b)]. The observed suppression of $1/^{23}T_1T$ below $T_{\rm SDW}$ is a consequence of the opening of the spin {\em gap} in the magnetic excitation spectrum. A concomitant drop in the stretching exponent $\gamma$ to the value $\sim 0.55$ in all three samples [Fig.~\ref{fig2}(c)] is due to the magnetic excitation spectrum becoming more anisotropic below $T_{\rm SDW}$ than above it. However, $1/^{23}T_1T$ does not approach zero at the lowest temperatures. Instead, below $20$~K the $1/^{23}T_1T$ data level off, adopting a simple Korringa relation. This speaks for an extra spin-lattice relaxation channel controlled by the {\em gapless} particle-hole excitations, which remains active down to the lowest temperatures. Such two spin-lattice relaxation channels are proposed in the two-fluid description of magnetic excitations below $T_{\rm SDW}$ that gives\cite{Smerald_2009}
\begin{equation}
\frac{1}{^{23}T_1T}=C_{\rm inc}+C_{\rm coh}\Phi\left(\frac{k_BT}{\Delta_S}\right).
\label{T1low}
\end{equation}
Here $\Delta_S$ is a spin gap, $\Phi(x)=x^2{\rm Li}_1(e^{-1/x})+x^3{\rm Li}_2(e^{-1/x})$ and ${\rm Li}_n(z)$ is the $n^{\rm th}$ polylogarithm of $z$. The first term, $C_{\rm inc}$, describes the relaxation via the incoherent gapless particle-hole excitations, and is thus proportional to the square of the residual density of states at the Fermi level. The second term describes the relaxation via the coherent spin wave excitations with dispersion $\omega_{\bf k}=\sqrt{\Delta_S^2+({\bf v}\cdot{\bf k})^2}$ as found in neutron scattering experiments,\cite{Smerald_2009} where ${\bf v}$ is the spin wave velocity. This model perfectly fits our $1/^{23}T_1T$ data below $T_{\rm SDW}$ [Fig.~\ref{fig2}(b)]. The obtained spin gap $\Delta_S=170(15)$~K ($14.7$~meV) is nearly the same for all samples and thus independent of the doping level $x$. Its value is comparable to those obtained for other iron pnictides, e.g., for SrFe$_2$As$_2$.\cite{Smerald_2009} In contrast, the strength of $C_{\rm inc}$, and thus of the gapless part of the Fermi surface, varies strongly with $x$. It amounts to $1.6\cdot10^{-3}$~(sK)$^{-1}$, $13\cdot10^{-3}$~(sK)$^{-1}$, and $2.3\cdot10^{-3}$~(sK)$^{-1}$ for $x=1$, $0.9$ and $0.8$, respectively, thus being almost an order of magnitude larger for $x=0.9$ than for $x=1$ and $0.8$. We thus conclude, that at $T\ll T_{SDW}$, much bigger portion of the Fermi surface  remains gapless in the optimally doped Na$_{0.9}$FeAs in comparison to the other two compositions. Simultaneously, it is only Na$_{0.9}$FeAs where we observe a {\em bulk} superconductivity below $T_c\approx 12$~K seen as a diamagnetic response in the dc spin susceptibility [Fig.~\ref{fig2}(a)] and as a further suppression of $1/^{23}T_1T$ below $T_c$ [Fig.~\ref{fig2}(b)]. When crossing $T_c$, the $^{23}$Na lineshape [see the linewidth in Fig.~\ref{fig1}(c)] and the stretching exponent $\gamma$ [see Fig.~\ref{fig2}(c)] do not exhibit any change, which suggests that the SC order in Na$_{0.9}$FeAs below $T_c$ coexists with the SDW order {\em on atomic scale}. This statement is supported also by $^{23}T_1^{-1}$ following neither the conventional Bardeen-Cooper-Schrieffer (BCS) nor the $T^3$ dependence [Fig.~\ref{fig2}(b) inset], meaning that the bulk superconductivity in Na$_{0.9}$FeAs is highly unconventional.

\begin{figure}
\includegraphics[width=0.8\linewidth]{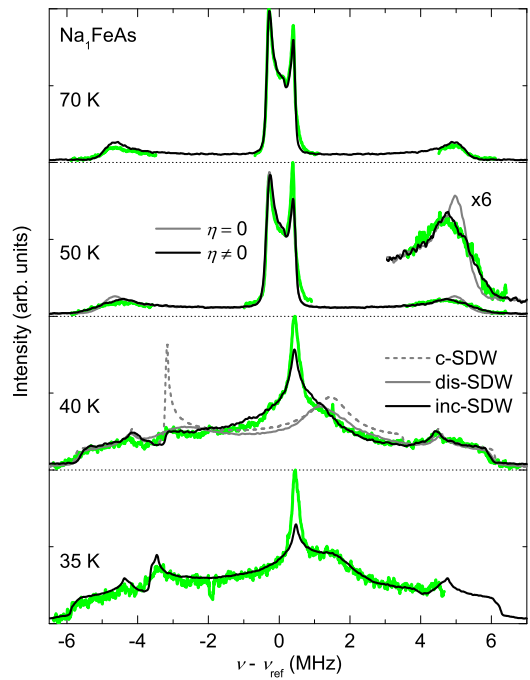}
\caption{$^{75}$As NMR spectra of powdered Na$_1$FeAs at $70$, $50$, $40$ and $35$~K (thick green lines). Comparison is made to the simulated spectra for a powdered sample with $\eta=0$ at $70$~K (thin black line), $\eta=0$ (thin gray line) and $0.1$ (thin black line) at $50$~K (with $6$-times magnified high-frequency singularity belonging to the satellite transition in the inset), and $\eta=0.15$ at $40$ and $35$~K. At $40$~K the following type of the SDW order is assumed: commensurate SDW (dashed thin gray line), disordered version of the commensurate SDW (solid thin gray line) and sinusoidal IC SDW (solid thin black line). At $35$~K a distorted sinusoidal IC SDW (solid thin black line) is assumed. The details are in the text.}
\label{fig3}
\end{figure}

The coexistence of the SC and the SDW orders on atomic scale is quite unusual and it is argued to be only possible when the SC state is unconventional.\cite{Fernandes_2010, Fernandes_2010_2} In a simplified case of spherical Fermi surfaces the coexistence is possible only when the SDW order is IC.\cite{Cvetkovic_2009,Vorontsov_2009,Vorontsov_2010} To disclose the nature of the SDW order in the studied samples we turn to the $^{75}$As NMR as $^{75}$As nuclei, being located in the FeAs layer, are much more sensitive to the local magnetic fields than $^{23}$Na nuclei. The following analysis applies to all three compositions, but we present it for the Na$_1$FeAs case, where the SDW order parameter is the biggest. As shown in Fig.~\ref{fig3}, a $^{75}$As frequency-swept NMR spectrum in the temperature range $70-300$~K is characteristic for a powdered sample with an axially symmetric EFG tensor ($\eta=0$), in accordance with the As site symmetry, and as found also in other iron pnictides.\cite{Jeglic_2009, Jeglic_2010} The spectrum consists of two symmetric singularities belonging to the satellite transitions ($-3/2\leftrightarrow -1/2$ and $1/2\leftrightarrow 3/2$) of the $I=3/2$ nuclear spin, located roughly at $\pm 5$~MHz with respect to the reference frequency, and of the narrow central part belonging to the central transition ($-1/2\leftrightarrow 1/2$), itself consisting of two singularities. The spectrum at $70$~K is simulated with the quadrupole frequency $\nu_Q$ distributed in the Gaussian manner in a narrow range of $0.24$~MHz around the value $10.2$~MHz (solid thin black line in Fig.~\ref{fig3}). Between $70$~K and $50$~K, still above $T_{\rm SDW}$, the spectrum changes. Namely, at $50$~K each satellite singularity is slightly split and thus less pronounced. As shown in Fig.~\ref{fig3} (and the corresponding inset), the spectrum can only be reproduced taking $\eta\approx 0.1$ (solid thin black line), meaning that the structural transition takes place at $T_S\approx 60$~K, comfortably above $T_{\rm SDW}$.

Below $T_{\rm SDW}$, the spectrum dramatically changes due to the appearance of internal magnetic fields $B_{\rm int}$ experienced by the $^{75}$As nuclei in the SDW state. Most notably, at $40$~K the low-frequency singularity of the central part shifts down to $\sim -3.2$~MHz and gets much less pronounced, while the high-frequency singularity of the central transition roughly keeps its position and gets much broader (Fig.~\ref{fig3}). In addition, the splitting of each satellite singularity increases, which can be accounted for by taking $\eta\approx 0.15$. A $^{75}$As lineshape simulation based on the commensurate SDW structure, suggested by neutron diffraction experiments on Na$_x$FeAs,\cite{Li_2009} fails to reproduce an experimental spectrum (dashed thin gray line in Fig.~\ref{fig3}). Namely, the high-frequency singularity of the central transition clearly misses the experimental position, while the low-frequency singularity of the central transition is much more pronounced than the experimental one. We thus test two alternative magnetic structures that were recently suggested. (i) For Ni-doped BaFe$_2$As$_2$ a commensurate SDW order with large local amplitude variations in the vicinity of the dopant was suggested.\cite{Dioguardi_2010} Although Na$_1$FeAs is stoichiometric, we speculate that a slight Na off-stoichiometry may lead to similar effects as Ni doping in BaFe$_2$As$_2$. However, introducing a Gaussian distribution of $B_{\rm int}$ at the As site as in Ref.~\onlinecite{Dioguardi_2010}, we again fail to reproduce an experimental spectrum (solid thin gray line in Fig.~\ref{fig3}). (ii) For lightly Co-doped BaFe$_2$As$_2$ an IC SDW order was suggested on the basis of $^{75}$As NMR results\cite{Laplace_2009, Ning_2009} and $^{57}$Fe M\"ossbauer spectroscopy results.\cite{Bonville_2010} Introducing an IC distribution of $B_{\rm int}$, $B_{\rm int}=B_{\rm int}^0\sin\alpha$ with $\alpha\in [0,2\pi]$ and $B_{\rm int}^0=0.45$~T, we manage to reproduce an experimental spectrum including both central and both satellite singularities very well (solid thin black line in Fig.~\ref{fig3}). Such a solution assumes internal local magnetic fields almost parallel to the crystallographic $c$-axis, and is thus consistent with the ordering vector $(1/2-\epsilon,0,1/2)$ for $\epsilon\rightarrow0$ (Ref.~\onlinecite{Laplace_2009}). We check this solution also against the low-temperature $^{23}$Na NMR lineshape. Its broadening with respect to the linewidth above $T_{\rm SDW}$ is governed by the dipolar interaction between the $^{23}$Na nuclei and the ordered magnetic moments, which enables us to extract the order parameter. Meanwhile, the very weak contact interaction is responsible for the $\sim 3$~kHz shift, which is much smaller than the low-temperature saturated linewidth. As shown in Fig.~\ref{fig2}(b) for Na$_1$FeAs, we are able to reproduce the saturated low-temperature $^{23}$Na NMR spectra very well by assuming an IC SDW amplitude of $0.28\mu_B$ [thin solid line in Fig.~\ref{fig2}(b)], while commensurate types of order do not fit [thin dashed and short dashed lines in Fig.~\ref{fig2}(b)]. Similar procedure leads to the IC SDW amplitudes of $0.04\mu_B$ and $0.13\mu_B$ for the $x=0.9$ and $0.8$ samples, respectively. The obtained order parameter for the Na$_1$FeAs sample is bigger than the one obtained in the muon-spin rotation~\cite{Parker_2009} and the neutron diffraction studies~\cite{Li_2009} but comparable to the one found in a recent NMR study~\cite{Yu_2010} and to those found in the members of 1111 family of iron pnictides. Finally, as shown in Fig.~\ref{fig3}, an experimental $^{75}$As spectrum below $40$~K again changes slightly. Namely, the spectrum at $35$~K acquires a hump at $\sim 1.5$~MHz, and the low-frequency singularity of the central transition gets slightly more pronounced, both features being characteristic of the commensurate SDW structure (see the discussion in the beginning of this paragraph). Indeed, a $^{75}$As spectrum at $35$~K can be reproduced reasonably well by an IC distribution of $B_{\rm int}$, $B_{\rm int}=B_{\rm int}^1\sin\alpha+B_{\rm int}^2\sin(3\alpha)$ with $\alpha\in [0,2\pi]$ and $B_{\rm int}^1=0.61$~T, $B_{\rm int}^2=0.18$~T (solid thin black line in Fig.~\ref{fig3}), similarly as in Ref.~\onlinecite{Bonville_2010}. Such a distribution departs from the pure sinusoidal distribution and approaches the square distribution, which corresponds to the SDW order with regions of uniform ordered moment separated by domain walls.

The observed incommensurability of the SDW wraps up the story of Na$_x$FeAs in the spirit of Refs.~\onlinecite{Cvetkovic_2009,Vorontsov_2009,Vorontsov_2010}, where the competition between the SC and the SDW order is studied theoretically. Namely, an incomplete nesting between electron and hole Fermi surfaces in Na$_x$FeAs is sufficient to stabilize the robust IC SDW state through opening of the spin gap only on a part of the Fermi surface. Large portion of the Fermi surface remains gapless below $T_{\rm SDW}$, and the larger this portion is the larger is the SC fraction below $T_c$ and the smaller is the saturated SDW order parameter. Apparently, the nesting is completely absent in Li$_1$FeAs as no SDW ordering is observed in this case despite strong AF fluctuations.\cite{Tapp_2008, Jeglic_2010} An incomplete or even absent nesting opens a possibility for the unconventional superconductivity following neither BCS nor the unconventional $T^3$ dependence of $T_1^{-1}$ below $T_c$, as shown in the inset of Fig.~\ref{fig2}(b) for Na$_{0.9}$FeAs and Li$_1$FeAs, respectively. The phase diagram emerging from the present work resembles the theoretically proposed one,\cite{Cvetkovic_2009,Vorontsov_2009,Vorontsov_2010} with an extra detail: a narrow SC dome is completely (not only partially) embraced by the IC SDW ordered state. This difference calls for further investigations, also in the light of the recent suggestion that the SDW order is not due to the nesting but rather due to the electronic band-structure reconstruction.\cite{He_2010}

\section{Conclusion}

Magnetic and SC properties of the Na$_x$FeAs series were probed by NMR. A pronounced variation of the spin-lattice relaxation with the level of carrier doping $x$ demonstrates the complicated multiband electronic structure around the Fermi level. This variation is particularly pronounced in the IC SDW state, where a significant portion of the Fermi surface remains gapless. Our results are directly compared to the recent theoretical treatments of the interplay between magnetism and superconductivity in iron pnictides, and thus promote the members of the 111 family as perfect model systems.

During the review process of this manuscript, a preprint appeared on the arXiv server\cite{Kitagawa_2010} reporting $^{23}$Na and $^{75}$As NMR experiments on a single crystal of Na$_1$FeAs implying similar conclusions about the nature of the SDW state in Na$_1$FeAs below $45$~K and similar value of the order parameter $\sim 0.3\mu_B$.

\begin{acknowledgements}
This work was supported by the ARRS project No. J1-2284-1. M. K. and P. J. acknowledge support from the Centre of Excellence EN$\rightarrow$FIST, Dunajska 156, SI-1000 Ljubljana, Slovenia. A. M. G. and B. L. acknowledge the NSF (Grant No. CHE-0616805) and the R. A. Welch Foundation (Grant No. E-1297) for support. The work in Houston is supported in part by U.S. Air Force Office of Scientific Research Award No. FA9550-09-1-0656, U.S. Department of Energy subcontract 4000086706 through Oak Ridge National Laboratory, U.S. Air Force Research Laboratory subcontract R15901 through Rice University, the T. L. L. Temple Foundation, the John J. and Rebecca Moores Endowment, and the State of Texas through the Texas Center for Superconductivity at the University of Houston; and at Lawrence Berkeley Laboratory by the Director, Office of Science, Office of Basic Energy Sciences, Division of Materials Sciences and Engineering of the U.S. Department of Energy under Contract No. DE-AC03-76SF00098.
\end{acknowledgements}


\begin{thebibliography}{99}
\bibitem{Paglione_2010} J. Paglione and R. L. Greene, Nature Phys. {\bf 6}, 645 (2010).
\bibitem{Johnston_2010} D. C. Johnston, Adv. Phys. {\bf 59}, 803 (2010).
\bibitem{Wilson_2010} J. A. Wilson, J. Phys.: Condens. Matter {\bf 22}, 203201 (2010).
\bibitem{Haule_2008} K. Haule, J. H. Shim and G. Kotliar, Phys. Rev. Lett. {\bf 100}, 226402 (2008).
\bibitem{Boeri_2008} L. Boeri, O. V. Dolgov and A. A. Golubov, Phys. Rev. Lett. {\bf 101}, 026403 (2008).
\bibitem{Lee_2006} P. A. Lee, N. Nagaosa, X.-G. Wen, Rev. Mod. Phys. {\bf 78}, 17 (2006).
\bibitem{Chu_2009} C. W. Chu, Nature Phys. {\bf 5}, 787 (2009).
\bibitem{Cs3Science} Y. Takabayashi, A. Y. Ganin, P. Jegli\v{c}, D. Ar\v{c}on, T. Takano, Y. Iwasa, Y. Ohishi, M. Takata, N. Takeshita, K. Prassides, and M. J. Rosseinsky, Science {\bf 323}, 1585 (2009).
\bibitem{Ganin2010} A. Y. Ganin, Y. Takabayashi, P. Jegli\v{c}, D. Ar\v{c}on, A. Poto\v{c}nik, P. J. Baker, Y. Ohishi, M. T. McDonald, M. D. Tzirakis, A. McLennan, G. R. Darling, M. Takata, M. J. Rosseinsky, and K. Prassides, Nature {\bf 466}, 221 (2010).
\bibitem{Luetkens_2009} H. Luetkens, H.-H. Klauss, M. Kraken, F. J. Litterst, T. Dellmann, R. Klingeler, C. Hess, R. Khasanov, A. Amato, C. Baines, M. Kosmala, O. J. Schumann, M. Braden, J. Hamann-Borrero, N. Leps, A. Kondrat, G. Behr, J. Werner, and B. B\"uchner, Nature Mater. {\bf 8}, 305 (2009).
\bibitem{Chen_2009} H. Chen, Y. Ren, Y. Qiu, W. Bao, R. H. Liu, G. Wu, T. Wu, Y. L. Xie, X. F. Wang, Q. Huang, and X. H. Chen, Europhys. Lett. {\bf 85}, 17006 (2009).
\bibitem{Ishida_2009} K. Ishida, Y. Nakai and H. Hosono, J. Phys. Soc. Jpn. {\bf 78}, 062001 (2009).
\bibitem{Ning_2010} F. L. Ning, K. Ahilan, T. Imai, A. S. Sefat, M. A. McGuire, B. C. Sales, D. Mandrus, P. Cheng, B. Shen, and H.-H Wen, Phys. Rev. Lett. {\bf 104}, 037001 (2010).
\bibitem{Tapp_2008} J. H. Tapp, Z. Tang, B. Lv, K. Sasmal, B. Lorenz, P. C. W. Chu, and A. M. Guloy, Phys. Rev. B {\bf 78}, 060505(R) (2008).
\bibitem{Chu_2009_2} C. W. Chu, F. Chen, M. Gooch, A. M. Guloy, B. Lorenz, B. Lv, K. Sasmal, Z. J. Tang, J. H. Tapp, and Y. Y. Xue, Physica C {\bf 469}, 326 (2009).
\bibitem{Sasmal_2009} K. Sasmal, B. Lv, Z. J. Tang, F. Chen, Y. Y. Xue, B. Lorenz, A. M. Guloy, and C. W. Chu, Phys. Rev. B {\bf 79}, 184516 (2009).
\bibitem{Parker_2009} D. R. Parker, M. J. Pitcher, P. J. Baker, I. Franke, T. Lancaster, S. J. Blundell, and S. J. Clarke, Chem. Commun. (Cambridge) {\bf 16} 2189 (2009).
\bibitem{Li_2009} S. Li, C. de la Cruz, Q. Huang, G. F. Chen, T.-L. Xia, J. L. Luo, N. L. Wang, and P. Dai, Phys. Rev. B {\bf 80}, 020504(R) (2009).
\bibitem{Yu_2010} W. Yu, L. Ma, J. Zhang, G. F. Chen, T.-L. Xia, S. Zhang, and Y. Hou, arXiv: 1004.3581 (2010).
\bibitem{Parker_2010} D. R. Parker, M. J. P. Smith, T. Lancaster, A. J. Steele, I. Franke, P. J. Baker, F. L. Pratt, M. J. Pitcher, S. J. Blundell, and S. J. Clarke, Phys. Rev. Lett. {\bf 104}, 057007 (2010).
\bibitem{Jishi_2010} R. A. Jishi and H. M. Alyahyaei, Adv. Condens. Matter Phys. 804343 (2010).
\bibitem{Abragam} A.~Abragam, {\sl Principles of Nuclear Magnetism} (Clarendon Press, Oxford, 1983).
\bibitem{Nakai} Y. Nakai, K. Ishida, Y. Kamihara, M. Hirano, and H. Hosono, J. Phys. Soc. Jpn. {\bf 77}, 073701 (2008).
\bibitem{Smerald_2009} A. Smerald and N. Shannon, Europhys. Lett. {\bf 92}, 47005 (2010).
\bibitem{Fernandes_2010} R. M. Fernandes, D. K. Pratt, W. Tian, J. Zarestky, A. Kreyssig, S. Nandi, M. G. Kim, A. Thaler, N. Ni, P. C. Canfield, R. J. McQueeney, J. Schmalian, and A. I. Goldman, Phys. Rev. B {\bf 81}, 140501(R) (2010).
\bibitem{Fernandes_2010_2} R. M. Fernandes and J. Schmalian, Phys. Rev. B {\bf 82}, 014521 (2010).
\bibitem{Cvetkovic_2009} V. Cvetkovic and Z. Tesanovic, Phys. Rev. B {\bf 80}, 024512 (2009).
\bibitem{Vorontsov_2009} A. B. Vorontsov, M. G. Vavilov and A. V. Chubukov, Phys. Rev. B {\bf 79}, 060508(R) (2009).
\bibitem{Vorontsov_2010} A. B. Vorontsov, M. G. Vavilov and A. V. Chubukov, Phys. Rev. B {\bf 81}, 174538 (2010).
\bibitem{Jeglic_2009} P. Jegli\v{c}, J.-W. G. Bos, A. Zorko, M. Brunelli, K. Koch, H. Rosner, S. Margadonna, and D. Ar\v{c}on, Phys. Rev. B {\bf 79}, 094515 (2009).
\bibitem{Jeglic_2010} P. Jegli\v{c}, A. Poto\v{c}nik, M. Klanj\v{s}ek, M. Bobnar, M. Jagodi\v{c}, K. Koch, H. Rosner, S. Margadonna, B. Lv, A. M. Guloy, and D. Ar\v{c}on, Phys. Rev. B {\bf 81}, 140511(R) (2010).
\bibitem{Dioguardi_2010} A. P. Dioguardi, N. apRoberts-Warren, A. C. Shockley, S. L. Bud�ko, N. Ni, P. C. Canfield, and N. J. Curro, Phys. Rev. B {\bf 82}, 140411(R) (2010).
\bibitem{Laplace_2009} Y. Laplace, J. Bobroff, F. Rullier-Albenque, D. Colson, and A. Forget, Phys. Rev. B {\bf 80}, 140501(R) (2009).
\bibitem{Ning_2009} F. L. Ning, K. Ahilan, T. Imai, A. S. Sefat, R. Jin, M. A. McGuire, B. C. Sales, and D. Mandrus, Phys. Rev. B {\bf 79}, 140506(R) (2009).
\bibitem{Bonville_2010} P. Bonville, F. Rullier-Albenque, D. Colson, and A. Forget, Europhys. Lett. {\bf 89}, 67008 (2010).
\bibitem{He_2010} C. He, Y. Zhang, B. P. Xie, X. F. Wang, L. X. Yang, B. Zhou, F. Chen, M. Arita, K. Shimada, H. Namatame, M. Taniguchi, X. H. Chen, J. P. Hu, and D. L. Feng, Phys. Rev. Lett. {\bf 105}, 117002 (2010).
\bibitem{Kitagawa_2010} K. Kitagawa, Y. Mezaki, K. Matsubayashi, Y. Uwatoko, and M. Takigawa, arXiv: 1011.1108 (2010).
\end{thebibliography}
\end{document}